\documentstyle[prl,aps,epsfig,twocolumn]{revtex}

\begin{document}
\draft

\twocolumn[\hsize\textwidth\columnwidth\hsize\csname@twocolumnfalse\endcsname
\title{Transient dynamics of Light propagation in EIT medium and hidden symmetry
of  multi-bit quantum  memory }
\author{Y. Li $^{1}$, X. F. Liu $^{1,2}$ and C. P. Sun $^{1,a,b}$}
\address{$^{1}$Institute of Theoretical Physics, The Chinese Academy of
Science, Beijing, 100080, China\\
$^{2}$Department of Mathematics, Peking University, Beijing,
China\\}

\date{\today}
\maketitle

\begin{abstract}
We investigate the transient phenomenon or property of the
propagation of an optical probe field in  a medium consisting of
many $\Lambda$-type three-level atoms coupled to this  probe field
and an classical driven field. We observe a hidden symmetry and
obtain an exact solution for this  light propagation problem by
means of the spectral generating method. This solution enlightens
us to propose a practical protocol implementing the quantum memory
robust for quantum decoherence in a crystal. As an transient
dynamic process this solution  also manifests  an exotic result
that a wave-packet of light will split into three packets
propagating at different group velocities. It is argued that
 "super-luminal group velocity" and "sub-luminal group
velocity" can be observed simultaneously in the same system. This
interesting phenomenon is expected to be demonstrated
experimentally.
\end{abstract}

]


\section{Introduction}

The recent experiments \cite{sub,super} have demonstrated many exotic
natures of light pulse propagation in the solid and gas systems with
electromagnetically induced transparency (EIT) \cite{EIT}, such as the
phenomena of "super-luminal group velocity"\ and "sub-luminal group
velocity". Physically they reflect the quantum coherence effects to
correlate the quantum fluctuations \cite{scully1}. The concepts of atomic
coherence and interference have also been applied to lasing without
inversion \cite{LWI} and the enhancement of linear or nonlinear
susceptibilities \cite{scully2,harris}. At present the sub-luminal phenomena
including the light stopping in an atomic medium have been extensively
studied from various points of view \cite{You}. The recent experiments have
also shown the ultra-slow group velocity of light in the solid state
systems, such as 3-mm thick crystal of $Y_{2}SiO_{5}$ \cite{exp-solid}. Most
recently, it is found that when a few rubidium atoms are loaded into high-Q
optical micro-cavity, the "super-luminal"\ and "sub-luminal"\ phenomena can
be observed in a refined version of the above\ mentioned experiments \cite%
{deng}.

With possible applications in quantum information field developed in the
past ten years \cite{q-inf}, some studies on the sub-luminal problem are
closely associated with an ideal and reversible transfer technique for the
quantum state between light and metastable collective states of matter \cite%
{store}. Theoretically, the basic idea is based on the control of light
propagation in a coherently driven 3-level atomic medium. The exciting fact
that the group velocity is adiabatically reduced to zero \cite{store} means
the possibility of proposing a more practical protocol to store and transfer
the quantum information of photons in the collective excitations in an
ensemble of $\Lambda $-type 3-level atoms \cite{lukin1,lukin2}. This may
open up an interesting prospective for quantum information processing. The
studies show that the excitations of matter coupled by light are more stable
under some circumstances and thus form the so-called "dark-state
polaritons". \ Physically speaking, this is the essence of the problem.
Extended to the multi-atom case, the above idea about adiabatic transfer of
the state of a single photon into that of an individual atom \cite{single}
provides a conceptually simplest approach for the implementation of the
quantum memory of photon information. Technically, this approach combines
the enhancement of the absorption cross section in multi-atom systems with
dissipation-free adiabatic passage techniques. In addition, this significant
investigation motivates the protocol for long distance quantum communication
\cite{L-qcommun} based on atomic ensemble.

Most recently we have deeply investigated the robustness of this kind of
quantum memory \cite{sun-you}. To avoid the spatial-motion induced
decoherence in an ensemble of free atoms, we have naturally proposed a
protocol that each $\Lambda $-type atom is fixed on a lattice site of a
crystal \cite{sun2}. As quantum memories robust for the spatial-motion
induced decoherence, the quasi-spin wave collective excitation of many $%
\Lambda $-type atoms forms a two-mode exciton system with a dynamic symmetry
(or a hidden symmetry) depicted by the semi-direct product algebra $SU(2)%
\overline{\otimes }h_{2}$ ($h_{2}$ is an algebra of two mode boson
operator). Physically, this hidden symmetry guarantees the stable spectral
structure of such dressed two-mode exciton system while its decoherence
implies a symmetry breaking. If one only considers quantum memory, one can
focus on the case of single mode light field. However, if one considers the
propagation of light pulse in this EIT medium, it is then necessary to
concern the multi-mode light field since a light pulse can be understood as
a superposition of the components of different frequencies. The aim of the
present and subsequent papers is to analyze the transient phenomenon or
property of the light propagation in this EIT medium, especially to
emphasize the role of the generalized hidden symmetry.

In this paper, our proposed system still consists of the quasi-spin wave
collective excitation of many $\Lambda $-type 3-level subsystems as in ref.%
\cite{sun2}. The meta-stable state still interacts with an exactly resonant
classical field, but the approximately resonant quantized light field, which
couples to the transition between the excited state and the relative ground
state, is no longer of single mode. We will prove that, in the large $N$
limit, one needs to introduce a pair of exciton operators for each mode of
quantized light field. These realize the infinite boson algebra local in the
mode space (or the frequency domain), but the collective quasi-spin
operators intertwining between the meta-stable states and the excited ones
generate a single global $SU(2)$ algebra. With the help of this hidden
symmetry and the corresponding spectral generating algebra method, we
exactly obtain the dressed spectra of the total system formed by the
infinite two-mode excitons coupling to a multi-mode quantized
electromagnetic field in the large $N$ limit. The exact solutions for the
eigen-states obtained in this way also include the multi-mode dark states
extrapolating from photon to one mode quasi-spin wave exciton. Actually,
these dressed states describe the polaritons only coupling a quasi-spin wave
exciton to a photon for some special case. With the EIT mechanism the
external classical field can be artificially manipulated to change the
dispersion properties of the quasi-spin wave excitonic medium dramatically
so that the quantized probing light can propagate in exotic ways. In this
way the quantum information can be coherently stored and transferred among
the multi-mode cavity photon and the multi-mode exciton system. Our studies
in this paper are substantially related to the hidden symmetry and its
breaking. The hidden symmetry leads to an exact class of solutions for the
light propagation problem, showing that a wave-packet of light will split
into three wave-packets which propagate at different group velocities,
namely, the "super-luminal group velocity", the "sub-luminal group
velocity"\ and the usual light group velocity.

\section{Collective Excitation of Three-Level Medium}

We consider the transient process for the weak light propagation in a medium
consisting of many three-level subsystems. It can be an ensemble of $N$ free
$\Lambda $-type atoms, or a crystal with $N$ lattice sites attached by $N$ $%
\Lambda $-type subspaces. In recent years, the similar exciton system in a
crystal slab with spatially fixed "two-level atoms" has been extensively
discussed with the emphasis on fluorescence process and relevant quantum
decoherence problem \cite{sun-liu},\cite{j-s}.

As shown in Fig. 1, the $\Lambda $-type subsystem possesses an excited state
$|a\rangle $, a ground state $|b\rangle $ and a meta-stable state $|c\rangle
$. The transition frequency $\omega _{ac}$ from $|a\rangle $ to $|c\rangle $
of each atom is resonantly driven by a classical field of Rabi-frequency $%
\Omega $. The transition frequency $\omega _{ab}$ from $|a\rangle $ to $%
|b\rangle $ is coupled to a multi-mode quantized field with the annihilation
operator $a_{k}$, and the coupling constant $g_{k}$ for the optical mode of
wave vectors ${\bf k}$. Strictly speaking, a multi-mode field can't be
resonantly coupled to an atomic transition since it is a superposition of
many components of different frequencies. We should emphasize here that we
only consider the case that the quantized field is a Gaussian wave packet in
the frequency domain and the center frequency $\omega _{0}$ just equals $%
\omega _{ab}$. When the frequency width of the wave packet is very small
compared with $\omega _{0}$ ($\Delta \omega \ll \omega _{0}$), from the
viewpoint of approximation it is\ reasonable to assume that the multi-mode
quantized field couples resonantly to the atomic transition from $|a\rangle $
to $|b\rangle $. Therefore, the interaction Hamiltonian of total system
reads
\begin{eqnarray}
H &=&\sum_{{\bf j}=1}^{N}\sum_{k}g_{k}a_{k}\exp (i{\bf k}\cdot {\bf r}%
_{j})\sigma _{ab}^{{\bf j}}  \nonumber \\
&&+\Omega \sum_{j=1}^{N}\exp (i{\bf q}\cdot {\bf r}_{j})\sigma _{ac}^{{\bf j}%
}+h.c.,  \label{1}
\end{eqnarray}%
where ${\bf r}_{j}$ $(j=1,2,\cdots ,N)$ denotes the position of the ${\bf j}%
^{th}$ subsystem, $N$ the total atomic number, ${\bf k}$ the wave vector of
quantized light of ${\bf k}^{th}$ mode and ${\bf q}$ the wave vector of the
classical light field. The flip operators $\sigma _{\alpha \beta }^{{\bf j}%
}=|\alpha \rangle _{{\bf jj}}\langle \beta |$ ($\alpha ,\beta =a,b,c$) for $%
\alpha \neq \beta $ define the quasi-spin between the given levels $\alpha $
and $\beta $. The coupling constant $g_{k}=-\wp \sqrt{\frac{kc}{2\hbar
\epsilon V}}$ depends on the matrix element $\wp $ of the electric dipole
moment between $|a\rangle $ and $|b\rangle $. For simplicity, $g_{k}$ and $%
\Omega $ are considered as real without loss of generality. As a matter of
fact, we need not require the resonances if we only consider the light
propagation in such EIT resonance \cite{l-z-s} \cite{lukin-rmp}.

%
\begin{figure}[h]
\begin{center}
\includegraphics[width=6cm,height=7cm]{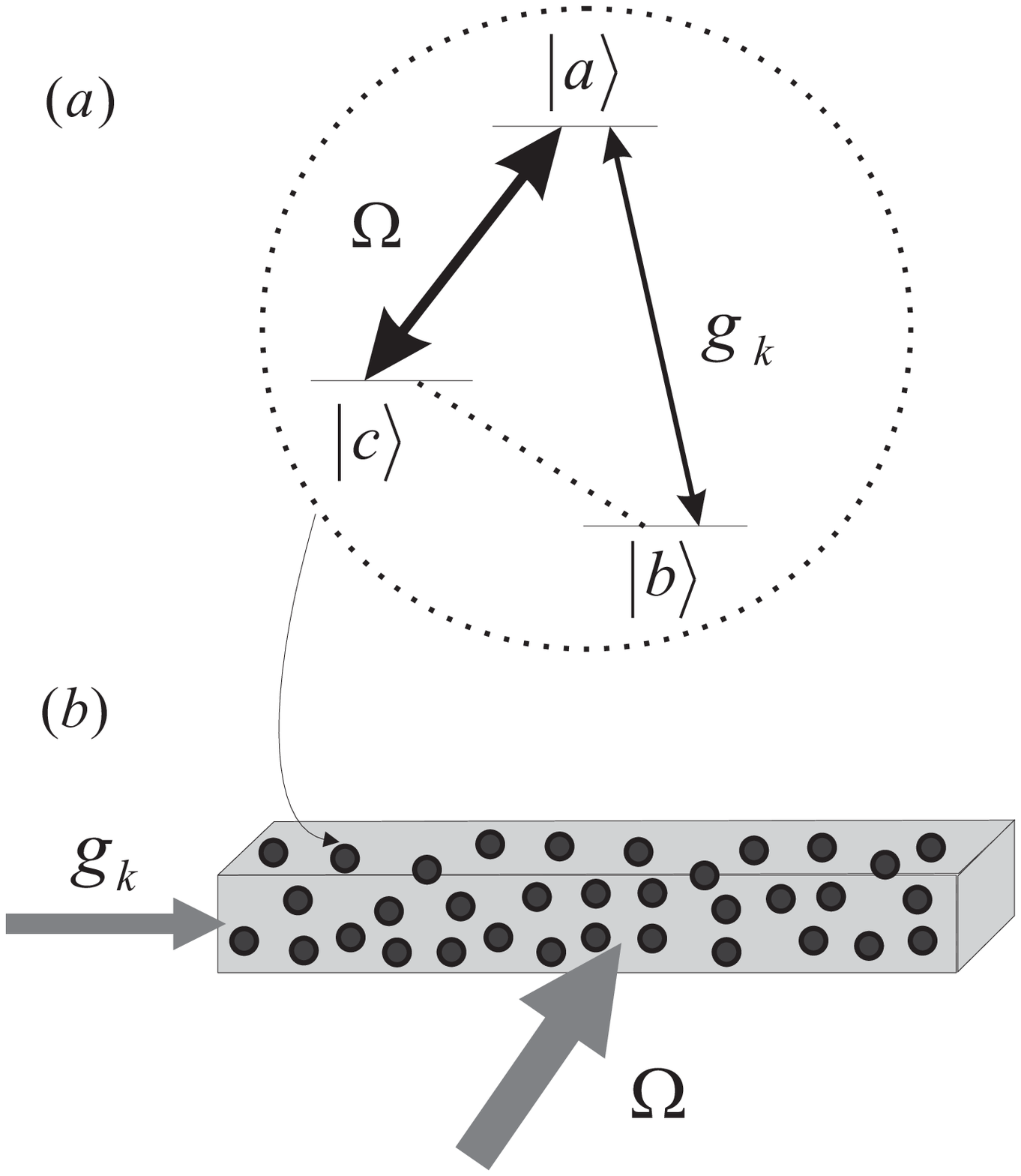}
\end{center}
\caption{Configuration of the quantum memory with $\Lambda $-type atoms. (a)
attached on lattice sites of crystal. (b) resonantly coupled to a control
classical field and a quantized probe field. }
\end{figure}

Generalizing the definition of the excitation operators of single optical
mode in the Ref. \cite{sun2}, we introduce a class of collective excitation
operators of optical multi-mode
\begin{equation}
A_{k}=\frac{1}{\sqrt{N}}\sum_{{\bf j}=1}^{N}e^{-i{\bf k}\cdot {\bf r}%
_{j}}\sigma _{ba}^{{\bf j}}
\end{equation}%
with respect to the transition from $|b\rangle $ to $|a\rangle $ for each
wave vector ${\bf k}$.{\bf \ }Correspondingly, the collective virtual
transition from $|b\rangle $ to $|c\rangle $ can be described by another
class of collective operators
\begin{equation}
C_{k}=\frac{1}{\sqrt{N}}\sum_{{\bf j}=1}^{N}e^{-i{\bf Q(k)}\cdot {\bf r}%
_{j}}\sigma _{bc}^{{\bf j}},
\end{equation}%
where ${\bf Q(k)}={\bf k}-{\bf q}$ means the momentum conservation in the
virtual process of collective transition from $|b\rangle $ to $|c\rangle $.
These collective operators create the general collective excitations
\[
|{\bf m},{\bf n}\rangle =%
\mathop{\displaystyle\prod}%
_{k}(\frac{1}{\sqrt{m_{k}!n_{k}!}}A_{k}^{\dagger m_{k}}C_{k}^{\dagger
n_{k}})|{\bf b}\rangle
\]%
defined for the set of multi-indices
\[
{\bf m=(}m_{1,}m_{2,}...),{\bf n=(}n_{1,}n_{2,}...),
\]%
and the collective ground state
\[
|{\bf b}\rangle =|b,b,\cdots ,b\rangle
\]%
with all $N$ atoms staying in the same single particle ground state $%
|b\rangle $. For example, the single particle excitations $|1_{a}\rangle
_{k}\equiv A_{k}^{\dagger }|{\bf b}\rangle $ and $|1_{c}\rangle _{k}\equiv
C_{k}^{\dagger }|{\bf b}\rangle $. These excitations are easy to understand.
Indeed, it is obvious that, from the ground state $|{\bf b}\rangle $, the
first order perturbation of the interaction creates the so-called one
exciton quasi-spin wave state
\begin{eqnarray}
|1_{a}\rangle _{k} &=&\frac{1}{\sqrt{N}}\sum_{{\bf j}=1}^{N}e^{i{\bf k}\cdot
{\bf r}_{j}}|b,b,\cdots ,\stackrel{{\bf j-}th}{\overbrace{a}},\cdots
,b\rangle ,  \nonumber \\
|1_{c}\rangle _{k} &=&\frac{1}{\sqrt{N}}\sum_{j=1}^{N}e^{i{\bf Q(k)}\cdot
{\bf r}_{j}}|b,b,\cdots ,\stackrel{{\bf j-}th}{\overbrace{c}},\cdots
,b\rangle .
\end{eqnarray}%
\bigskip

Physically, in the large $N$ limit under the low excitation condition, that
is, there are only a few atoms occupying the states $|a\rangle $ or $%
|c\rangle $ \cite{q-defor} and the population of state $|b\rangle $ is
approximately $N\rightarrow \infty $, the above two classes of quasi-spin
wave excitations behave as two classes of bosons since they have the
following bosonic commutation relations
\begin{eqnarray}
\lbrack A_{k},A_{k^{\prime }}^{\dagger }] &=&\delta _{k,k^{\prime }},
\nonumber \\
\lbrack C_{k},C_{k^{\prime }}^{\dagger }] &=&\delta _{k,k^{\prime }}, \\
\lbrack A_{k},C_{k^{\prime }}^{\dagger }] &=&0.  \nonumber
\end{eqnarray}%
To prove the above basic commutation relations, we first calculate
\begin{eqnarray*}
\lbrack A_{k},A_{k^{\prime }}^{\dagger }] &=&=\frac{1}{N}\sum_{{\bf j}%
=1}^{N}e^{-i({\bf k}^{\prime }-{\bf k)}\cdot {\bf r}_{j}}(\sigma _{bb}^{{\bf %
j}}-\sigma _{aa}^{{\bf j}}) \\
&\simeq &\frac{1}{N}\sum_{{\bf j}=1}^{N}e^{-i({\bf k}^{\prime }-{\bf k)}%
\cdot {\bf r}_{j}}.
\end{eqnarray*}%
Here we have considered that
\begin{equation}
\sum_{{\bf j}=1}^{N}e^{-i({\bf k}^{\prime }-{\bf k)}\cdot {\bf r}%
_{j}}(\sigma _{bb}^{{\bf j}}-\sigma _{aa}^{{\bf j}})\approx \sum_{{\bf j}%
=1}^{N}e^{-i({\bf k}^{\prime }-{\bf k)}\cdot {\bf r}_{j}}\sigma _{bb}^{{\bf j%
}}
\end{equation}%
while
\begin{equation}
\sum_{{\bf j}=1}^{N}e^{-i({\bf k}^{\prime }-{\bf k)}\cdot {\bf r}_{j}}\sigma
_{aa}^{{\bf j}}\approx 0
\end{equation}%
in the case of low excitation. For a crystal with regular lattice structure,
according to the theory of solid state, we should have
\begin{equation}
\frac{1}{N}\sum_{{\bf j}=1}^{N}e^{-i({\bf k}^{\prime }-{\bf k)}\cdot {\bf r}%
_{j}}=\delta _{{\bf k,k}^{\prime }}.
\end{equation}%
In fact, this result can be extended to the case of generic medium. For a
medium with a scale $L$ much larger than $\frac{2\pi }{\left\vert {\bf k}%
\right\vert },\frac{2\pi }{\left\vert {\bf k}^{\prime }\right\vert }$ (this
is valid for normal case), the summation over ${\bf r}_{j}$ can be replaced
by integral and it follows that
\begin{equation}
\frac{1}{N}\sum_{{\bf j}=1}^{N}e^{-i({\bf k}^{\prime }-{\bf k)}\cdot {\bf r}%
_{j}}=\frac{1}{V_{me}}\int_{V_{me}}e^{-i({\bf k}^{\prime }-{\bf k)}\cdot
{\bf r}}d^{3}{\bf r}\simeq \delta _{{\bf k,k}^{\prime }}
\end{equation}%
since the medium volume $V_{me}\rightarrow \infty $. The other commutation
relations can be proved in a similar way.\bigskip

\section{Parallelism Quantum Memory with Hidden Symmetry}

As argued above physically the operators $A_{k}$ and $C_{k}$ depict the
collective-excitation processes of bosonic type. The importance of \ these
operators lies in that they define an invariant subspace
\begin{equation}
V_{C}:span\{|{\bf m},{\bf n}\rangle |{\bf m=(}m_{1,}m_{2}...),{\bf n=(}%
n_{1,}n_{2,}...)\}
\end{equation}%
for the interaction Hamiltonian (\ref{1}). This\ means that driven by this
Hamiltonian, any collective state from $V_{C}$ involves to a new collective
state still in $V_{C}$. Thus we can use the collective excitations as basic
blocks to describe the quantum dynamic process with a hidden symmetry.

Let us introduce the following additional collective operators concerning
the transition from $|c\rangle $ to $|a\rangle $
\begin{equation}
T_{-}=\sum_{{\bf j}=1}^{N}e^{-i{\bf q}\cdot {\bf r}_{j}}\sigma _{ca}^{{\bf j}%
},T_{+}=(T_{-})^{\dagger }
\end{equation}%
which is resonantly driven by a classic light. Together with the third
collective operator
\[
T_{3}=\sum_{{\bf j}=1}^{N}(\sigma _{aa}^{{\bf j}}-\sigma _{cc}^{{\bf j}})/2,
\]%
they generate $SU(2)$ algebra globally. The commutation relations \ between $%
SU(2)$ and the collective operators $A_{k}$ and $C_{k}$ are easy to
calculate. The non-vanishing commutators are as follows:

\begin{eqnarray}
\lbrack T_{-},C_{k}] &=&-A_{k},[T_{+},A_{k}]=-C_{k},  \nonumber \\
\left[ T_{+},C_{k}^{\dagger }\right] &=&A_{k}^{\dagger },\left[
T_{-},A_{k}^{\dagger }\right] =C_{k}^{\dagger }.
\end{eqnarray}%
The above close commutation relations and the expression of the interaction
Hamiltonian in terms of these excitation operators:
\begin{equation}
H=\sum_{k}g_{k}\sqrt{N}a_{k}A_{k}^{\dagger }+\Omega T_{+}+h.c.
\end{equation}%
show that there is a dynamic "group"\ (algebra) $G_{d}$ for the light
excited system, which is generated by $A_{k}$, $C_{k}$, $A_{k}^{\dagger }$, $%
C_{k}^{\dagger }$ ($k=1,2,...$), $T_{\pm }$ and $T_{3}$. Using $\Xi $ to
denote the Heisenberg-Weyl algebra generated by $A_{k}$, $C_{k}$, $%
A_{k}^{\dagger }$\ and $C_{k}^{\dagger }$ ($k=1,2,...$), we observe that the
dynamic "group"\ $G_{d}=SU(2)\stackrel{-}{\otimes }$ $\Xi $ is a semi-direct
product of $SU(2)$ and $\Xi $ because
\begin{equation}
\lbrack SU(2),\Xi ]\subset \Xi .
\end{equation}%
Actually the above dynamic symmetry of $G_{d}$ is a straightforward
generalization of the symmetry of the single quantized optical mode. With
this \ symmetry the Hamiltonian $H$ can be diagonalized in an elegant way by
means of the spectrum generating algebra method \cite{algebra} .

To this end we calculate the commutators of $H$ with the generators of $%
G_{d} $ and light field operators respectively:

\begin{eqnarray}
\lbrack C_{k},H] &=&\Omega A_{k},[A_{k},H]=\Omega C_{k},  \nonumber \\
\lbrack a_{k},H] &=&g_{k}\sqrt{N}A_{k}.  \label{15}
\end{eqnarray}%
It follows that the dark-state polariton operators
\begin{equation}
D_{k}=a_{k}\cos \theta _{k}-C_{k}\sin \theta _{k}
\end{equation}%
commute with the Hamiltonian $H$ for the $\theta _{k}$ satisfying
\begin{equation}
\tan \theta _{k}=\frac{g_{k}\sqrt{N}}{\Omega }.
\end{equation}%
It is obvious that the dark-state polariton operators satisfy the bosonic
commutation relations and define new dressed excitations mixing the
electromagnetic field and collective excitations of quasi spin wave.
Especially these new excitations are stable since
\begin{equation}
\lbrack D_{k},H]=0.
\end{equation}%
For the construction of the complete collective space dressed by the
quantized light fields, another ingredient is the bright-state polariton
operators
\begin{equation}
B_{k}=a_{k}\sin \theta _{k}+C_{k}\cos \theta _{k}
\end{equation}%
satisfying
\[
\lbrack D_{k},B_{k}]=0.
\]%
It also extrapolates from the light field of $a_{k}$ to the exciton of $%
C_{k} $ when one adiabatically changes $\theta _{k}$ from $\frac{\pi }{2}$
to zero. Evidently, the product state $|{\bf 0}\rangle =|{\bf b}\rangle
\otimes |0\rangle _{l}$ is an eigen-state of $H$ with zero eigen-value
where\ $|0\rangle _{l}$ is the vacuum of the electromagnetic field. So we
can construct a degenerate class of zero eigen-value-states or dark states
\begin{equation}
|d({\bf n)}\rangle \equiv |d({\bf n,}t{\bf )}\rangle =%
\mathop{\displaystyle\prod}%
_{k}\frac{1}{\sqrt{n_{k}!}}D_{k}^{\dagger n_{k}}|{\bf 0}\rangle ,
\end{equation}%
where the adiabatic time-dependence originates from the change of
Rabi-frequency $\Omega $ for the artificially-controlled classical field.
The fact that dark states are cancelled by $H$ means they can trap the
electromagnetic radiation from the excited states. Physically this is due to
quantum interference cancelling.

With the help of the above hidden symmetry we can also construct the energy
spectra with non-zero eigenvalues by using the bright-state polariton
operators. We start from the derived commutation relations from the above
equation (\ref{15})
\begin{eqnarray}
\lbrack H,B_{k}^{\dagger }] &=&\Theta _{k}A_{k}^{\dagger }, \\
\lbrack H,A_{k}^{\dagger }] &=&\Theta _{k}B_{k}^{\dagger },
\end{eqnarray}%
where
\begin{equation}
\Theta _{k}=\sqrt{g_{k}^{2}N+\Omega ^{2}}.
\end{equation}%
This prompts us to consider the two commuting quasi-boson operators
\begin{equation}
Q_{k\pm }=\frac{1}{\sqrt{2}}(A_{k}\pm B_{k}).
\end{equation}%
As we have the commutation relations
\begin{equation}
\lbrack H,Q_{k\pm }^{\dagger }]=\pm \Theta _{k}Q_{k\pm }^{\dagger },
\end{equation}%
the operators $Q_{k\pm }$ can serve as ladder operators to generate the
spectra of $H.$ In fact they produce the dressed states of the multi-mode
exciton system

\begin{equation}
|e({\bf m,s,n)}\rangle =%
\mathop{\displaystyle\prod}
_{k}\frac{1}{\sqrt{m_{k}!s_{k}!}}Q_{k+}^{\dagger m_{k}}Q_{k-}^{\dagger
s_{k}}|d({\bf n)}\rangle
\end{equation}
as the eigen-states of $H$ with the eigen-values
\begin{equation}
E({\bf m,s)}=\sum_{k}(m_{k}-s_{k})\Theta _{k},
\end{equation}
of infinite degeneracy degrees.

Now we consider the relation between the degeneracy of spectrum and the
hidden symmetry. We notice that the infinite degeneracy of $|e({\bf 0,0,n)}%
\rangle =|d({\bf n)}\rangle $ results from the generalized translation
symmetry described by
\begin{eqnarray}
T_{k} &=&\exp (\alpha _{k}D_{k}^{\dagger }-\alpha _{k}^{\ast }D_{k})
\nonumber \\
&=&{\cal D}_{a_{k}}(\alpha _{k}\cos \theta _{k}){\cal D}_{C_{k}}(-\alpha
_{k}\sin \theta _{k}),
\end{eqnarray}%
where
\[
{\cal D}_{a}(\alpha )=\exp (\alpha a^{\dagger }-\alpha ^{\ast }a)
\]%
is the usual coherent state generator. In fact, for the original Hamiltonian
(\ref{1}), the unitary transformation ${\cal D}_{a_{k}}(\alpha _{k}\cos
\theta _{k})$ changes the term $g_{k}\sqrt{N}a_{k}A_{k}^{\dagger }$ by $g_{k}%
\sqrt{N}A_{k}^{\dagger }\alpha _{k}\cos \theta _{k}$ while ${\cal D}%
_{C_{k}}(-\alpha _{k}\sin \theta _{k})$ changes the terms $\Omega T_{+}$ by $%
-g_{k}\sqrt{N}A_{k}^{\dagger }\alpha _{k}\sin \theta _{k}$ since
\[
{\cal D}_{a_{k}}^{-1}(\alpha _{k}\cos \theta _{k})a_{k}{\cal D}%
_{a_{k}}(\alpha _{k}\cos \theta _{k})=a_{k}+\alpha _{k}\cos \theta _{k},
\]%
\[
{\cal D}_{C_{k}}^{-1}(-\alpha _{k}\sin \theta _{k})T_{+}{\cal D}%
_{C_{k}}(-\alpha _{k}\sin \theta _{k})=T_{+}-A_{k}^{\dagger }\alpha _{k}\sin
\theta _{k}.
\]%
These two changes cancel each other. Besides this degeneracy for different $%
{\bf n}${\bf ,} due to the relation
\begin{equation}
\lbrack H,Q_{k+}^{\dagger }Q_{k-}^{\dagger }]=0,
\end{equation}%
there are a large additional class of dark states of zero-eigen-value. They
can be constructed by acting $Q_{k+}^{\dagger }Q_{k-}^{\dagger }$\
repeatedly on $|d({\bf n)}\rangle $.

The above arguments suggest that, compared with the single mode exciton
system the dressed multi-mode exciton system generated by $A_{k}^{\dagger
},B_{k}^{\dagger }$ and $a_{k}^{\dagger }$ can serve as a quantum memory
with more advantages. It is very interesting that all the parameters $\theta
_{k}$ for different modes of $k$ can be well controlled to change from zero
to infinity by a unique extremely slow adiabatic parameter $\Omega =\Omega
(t)$ varying from a very large value to zero. As usual, the quantum
information of photons is described by the superposition state
\begin{equation}
|L(0)\rangle =\sum_{{\bf n}}c_{{\bf n}}\,|{\bf n}\rangle _{l}\equiv \sum_{%
{\bf n}}c_{{\bf n}}%
\mathop{\displaystyle\prod}%
_{k}\frac{1}{\sqrt{n_{k}!}}a_{k}^{\dagger n_{k}}|0\rangle _{l}
\end{equation}%
where $|{\bf n}\rangle _{l}$ is a multi-mode Fock state for the light field.
In the case $\Omega (0)\rightarrow \infty $, whatever the values of the
parameters are we have $\sin \theta _{k}=0$ so $|d({\bf n,0)}\rangle =|{\bf b%
}\rangle \otimes |{\bf n}\rangle _{l}$ and the initial state of total system
can be expressed as a superposition of the adiabatic dark states $|d({\bf %
n,0)}\rangle $ at time $t=0:$
\begin{equation}
|S(0)\rangle =|{\bf b}\rangle \otimes \sum_{{\bf n}}c_{{\bf n}}\,|{\bf n}%
\rangle _{l}=\sum_{{\bf n}}c_{{\bf n}}\,|d({\bf n,0)}\rangle
\end{equation}%
By adiabatically changing $\Omega $ to zero at time $T$, each component in
the above superposition varies in this way:
\[
|d({\bf n,0)}\rangle \rightarrow |d({\bf n,T)}\rangle =|{\bf n}\rangle
_{C}\otimes |0\rangle _{l}.
\]%
Thus one extrapolates $|S(0)\rangle $ to a pure multi-exciton state
\begin{equation}
|S(T)\rangle =\sum_{{\bf n}}c_{{\bf n}}\,|{\bf n}\rangle _{C}\otimes
|0\rangle _{l},
\end{equation}%
where

\begin{equation}
|{\bf n}\rangle _{C}=%
\mathop{\displaystyle\prod}%
_{k}\frac{1}{\sqrt{n_{k}!}}C_{k}^{\dagger n_{k}}|{\bf b}\rangle
\end{equation}%
is the multi mode exciton state describing the collective excitation.

\bigskip

It is worth pointing out that, the above observation based on the dynamic
symmetry does not depend on the adiabaticallity of manipulation on the
external parameters at all. In our example, because the dark states don't
contain any atomic excited state $|a\rangle $, the spontaneous emission is
then forbidden even for a non-adiabatic manipulation. By the way\ we also
point out that the quantum parallelism of the multi-mode exciton quantum
memory gives rise to a very convenient encoding process. For a decimal
system number $n$ there is a unique binary representation:
\begin{equation}
n=\sum_{i=0}^{L-1}n_{i}2^{i}
\end{equation}%
where $n_{i}=0,1$. Then\ we can encode it in a special multi-mode Fock state

\begin{equation}
|{\bf n}\rangle _{l}=|n_{0,}n_{1,}n_{2,}...,n_{L-1}\rangle
\end{equation}%
of light field. The multi-mode Fock state $|{\bf n}\rangle _{l}$ can be
represented by an array of the many-photon qubits, with at most one photon
in each photon qubit. It is easy to prepare such photon state in flying
(photon) qubit firstly and then store the decimal system number $n$ in the
above multi-mode exciton quantum memory as a single component many-exciton
state $|{\bf n}\rangle _{C}$.

\section{Transient Dynamics of Light Propagation}

In the previous sections, making full use of the hidden symmetry, we not
only construct two classes of dark states and their generated spectra of
non-zero eigenvalues, but also demonstrate that coherent optical information
of multi-qubit can be stored in a medium with collective effect in EIT. The
storing and reading-out processes are controlled by stimulated photon
transfer between the classical and quantized light field. In this section we
will consider the transient dynamic process for the interaction between
light fields and the EIT medium. Some exotic natures of light propagation in
such a medium will be investigated with the help of the dynamic symmetry
analysis.

The evolution of the Heisenberg operators corresponding to the optical field
and the collective excitations can be described as
\begin{eqnarray}
\stackrel{.}{a}_{k} &=&-i\sqrt{N}g_{k}A_{k},  \nonumber \\
\stackrel{.}{A}_{k} &=&-i\sqrt{N}g_{k}a_{k}-i\Omega C_{k}, \\
\stackrel{.}{C}_{k} &=&-i\Omega A_{k}  \nonumber
\end{eqnarray}%
for each light mode ${\bf k}$.{\bf \ }The above motion equations of $a_{k}$,
$A_{k}$ and $C_{k}$ with respect to a given optical mode ${\bf k}$ are
coupled together, but there is no coupling of mode ${\bf k}$ operators to
those of different optical mode ${\bf k}^{\prime }$. Due to the semi-direct
product property of the dynamic group $G=SU(2)\stackrel{-}{\otimes }\Xi
\otimes \Gamma ,$ the generators $T_{+}$, $T_{-}$, and $T_{3}$ of $SU(2)$
algebra do not occur in the above system of Heisenberg equations. Here $%
\Gamma $ is the Heisenberg-Weyl group generated by the creation and
annihilation operators $a_{k}^{\dagger }$ and $a_{k}$. It leads to the
evolution equations of the bright- and dark- state polariton operators
\begin{eqnarray}
\stackrel{.}{A}_{k} &=&-i\sqrt{g_{k}^{2}N+\Omega ^{2}}B_{k},  \nonumber \\
\stackrel{.}{B}_{k} &=&-i\sqrt{g_{k}^{2}N+\Omega ^{2}}A_{k}, \\
\stackrel{.}{D}_{k} &=&0  \nonumber
\end{eqnarray}%
in a straightforward way. The above equations manifest the basic features of
the dark states: decoupled from other\ states and stable in the time
evolution.

With the initial conditions determined by those for\ $C_{k}$ and $a_{k}$,
the exact solutions of $B_{k}$ and $D_{k}$ are obtained as follows:
\begin{equation}
D_{k}(t)=D_{k}(0)=a_{k}(0)\cos \theta _{k}-C_{k}(0)\sin \theta _{k},
\end{equation}%
\begin{equation}
B_{k}(t)={\bf O}_{1}e^{-it\Theta _{k}}+{\bf O}_{2}e^{it\Theta _{k}}],
\end{equation}%
\begin{eqnarray*}
{\bf O}_{1} &=&\frac{1}{2}(B_{k}(0)-A_{k}(0)) \\
{\bf O}_{2} &=&\frac{1}{2}(B_{k}(0)+A_{k}(0))
\end{eqnarray*}%
where $a_{k}(0),$ $A_{k}(0),$ $B_{k}(0)$ and $C_{k}(0)$ are the initial
Heisenberg operators, and $\Theta _{k}=\sqrt{g_{k}^{2}N+\Omega ^{2}}$ is the
light field dressed Rabi frequency that measures the effective coupling of
excitonic states to the external field.

The dark-state polariton operator $D_{k}(t)$ is indeed a time-independent
constant. That is, in the Heisenberg picture, the operator $D_{k}(t)$ always
retains the initial ${\bf k-}$mode dark-state polariton operator $D_{k}(0)$
for each ${\bf k}$ and the atomic system always keeps being an EIT medium.
Here for simplicity we assume that the quantized light field propagates
along the $x-$axis. Then, the positive frequency part of the quantized light
field
\begin{equation}
E^{+}(x,t)=\sum_{k}\sqrt{\frac{\hbar ck}{2\epsilon V}}a_{k}(t)e^{ikx-ikct}
\end{equation}%
can be expressed explicitly in terms of the mode operators of quantized
light field

\begin{eqnarray}
a_k(t) &=&D_k(t)\cos \theta _k+B_k(t)\sin \theta _k  \nonumber \\
&=&[a_k(0)\cos \theta _k-C_k(0)\sin \theta _k]\cos \theta _k \\
&&-({\bf O}_1e^{-it\Theta _k}+{\bf O}_2e^{it\Theta _k})\sin \theta _k,
\nonumber
\end{eqnarray}
\ which results from the expressions of bright- and dark- state polariton
operators given above.

With the above simple solution we can straightforwardly investigate how the
quantized light propagates in the EIT medium. We assume that the collective
state of the medium is initially in the ground state $|{\bf b\rangle }$ and
the initial state of the light field is a wave packet. The electromagnetic
field quantized, the wave packet of light can be depicted by the direct
product of many coherent states, or the multi-mode coherent state
\begin{equation}
|{\bf \alpha }\rangle =\prod_{k}\otimes |\alpha _{k}\rangle ,
\end{equation}%
where $\alpha _{k}=\exp [-f^{2}(k-k_{0})^{2}]$ (up to a normalized factor)
since the probe light is a Gaussian wave packet in frequency domain. This is
because the expectation value of the free quantized electromagnetic field
operator
\[
E^{+}(x,t)=\sum_{k}\sqrt{\frac{\hbar ck}{2\epsilon V}}a_{k}(t)e^{ikx-ikct}
\]%
resembles a classical wave packet as the superposition of infinite
components of different frequencies. In interaction with EIT medium, the
mean of $a_{k}(t)$ over the initial state $|\psi (0)\rangle =|{\bf b\rangle }%
\otimes |{\bf \alpha }\rangle $ is
\begin{eqnarray}
\langle a_{k}(t)\rangle &=&\langle a_{k}(0)\rangle \cos ^{2}\theta _{k}+
\nonumber \\
&&\sin \theta _{k}(\langle {\bf O}_{2}\rangle e^{-it\Theta _{k}}+\langle
{\bf O}_{1}\rangle e^{it\Theta _{k}})  \nonumber \\
&=&\langle a_{k}(0)\rangle (\cos ^{2}\theta _{k}+\sin ^{2}\theta _{k}\cos
\Theta _{k}t),
\end{eqnarray}%
where we have used
\begin{eqnarray*}
\langle A_{k}(0)\rangle &=&\langle C_{k}(0)\rangle =0, \\
\langle B_{k}(0)\rangle &=&\langle a_{k}(0)\rangle \sin \theta _{k}.
\end{eqnarray*}%
So the mean of $E^{+}(x,t)$ is decomposed into three parts:
\begin{eqnarray}
\langle E^{+}(x,t)\rangle &=&\sum_{k}\sqrt{\frac{\hbar ck}{2\epsilon V}}%
\langle a_{k}(0)\rangle \times  \nonumber \\
&&(\cos ^{2}\theta _{k}+\sin ^{2}\theta _{k}\cos \Theta _{k}t)e^{i(kx-kct)}
\nonumber \\
&=&E_{0}^{+}(x,t)+E_{+}^{+}(x,t)+E_{-}^{+}(x,t)
\end{eqnarray}%
where
\begin{eqnarray}
E_{+}^{+}(x,t) &=&\sum_{k}\sqrt{\frac{\hbar ck}{2\epsilon V}}\langle
a_{k}(0)\rangle \sin ^{2}\theta _{k}e^{i[k(x-ct)-\Theta _{k}t]},  \nonumber
\\
E_{0}^{+}(x,t) &=&\sum_{k}2\sqrt{\frac{\hbar ck}{2\epsilon V}}\langle
a_{k}(0)\rangle \cos ^{2}\theta _{k}e^{ik(x-ct)},  \label{45} \\
E_{-}^{+}(x,t) &=&\sum_{k}\sqrt{\frac{\hbar ck}{2\epsilon V}}\langle
a_{k}(0)\rangle \sin ^{2}\theta _{k}e^{i[k(x-ct)+\Theta _{k}t]}.  \nonumber
\end{eqnarray}%
It should be noticed that the light field dressed Rabi frequency $\Theta
_{k} $ modifies the dispersion relations of light propagation in the exciton
medium.

The three parts in the above decomposition of $\langle E^{+}(x,t)\rangle $
can be understood as three wave packets spreading in the coordinate space.
We can see this point by considering the initial quantized light field in
the frequency domain as a Gaussian wave packet with the center frequency $%
\omega _{0}$ ($=k_{0}c=\omega _{ab}$). For the first wave packet $%
E_{+}^{+}(x,t)$, its position center is determined by maximizing the
wave-vector-dependent phase:
\begin{equation}
\frac{\partial }{\partial k}(kct+\Theta _{k}t-kx)|_{k_{0}}=0.
\end{equation}%
This determines the group velocity of the wave packet $E_{+}^{+}(x,t)$ as
\begin{equation}
V_{g_{+}}=\frac{\partial x}{\partial t}|_{k_{0}}=c+\frac{\partial \Theta _{k}%
}{\partial k}|_{k_{0}}.
\end{equation}%
The term on the r.h.s. simply leads to a modification of the group velocity
\begin{equation}
V_{g_{+}}=c(1+\frac{\Omega }{2k_{0}c}\frac{n}{\sqrt{1+n}}),  \label{48}
\end{equation}%
where $n=\frac{g_{k_{0}}^{2}N}{\Omega ^{2}}$. This is one of the central
results of this paper. It is evident that the first wave packet $%
E_{+}^{+}(x,t)$ represents the "super-luminal" light propagation with the
group velocity larger than the light speed $c$. It is also emphasized that
the "super-luminal" light propagation can be understood according to the
classical theory of wave propagation in an anomalous dispersion medium. It
is the interference between different frequency components that results in
this rather counterintuitive effect. In this sense, we think that the
"super-luminal" light pulse propagation observed in experiments is not too
odd and there is no direct connection between this phenomenon and the
causality in relativity.

It is not surprising to see the usual group velocity $V_{g_{0}}=c$ for the
second wave packet $E_{0}^{+}(x,t)$. But for the third wave packet $%
E_{-}^{+}(x,t)$ we find the sub-luminal group velocity similarly:

\begin{eqnarray}
V_{g_{-}} &=&c-\frac{\partial \Theta _{k}}{\partial k}|_{k_{0}}  \nonumber \\
&=&c(1-\frac{\Omega }{2k_{0}c}\frac{n}{\sqrt{1+n}}).  \label{49}
\end{eqnarray}%
This sub-luminal group velocity phenomenon is expected\ to find applications
in quantum information processing. In fact, it is common sense to believe
that the system stopping and slowing light could be used to store quantum
information of photon qubits and in the storage time quantum information
processing may be possible if we can sufficiently reduce the dissipative
loss during the "reading" and "writing" operations.

We can also analytically integrate out $E_{0}^{+}(x,t)$, $E_{+}^{+}(x,t)$
and $E_{-}^{+}(x,t)$ and obtain an explicit depiction of the propagating
wave packet. Since $g_{k}=-\wp \sqrt{\frac{kc}{2\hbar \epsilon V}}$, it is
convenient to write $g_{k}^{2}N$ as $G^{2}k$ with $G=\wp \sqrt{\frac{cN}{%
2\hbar \epsilon V}}$. Replacing the summation over $k$ by integral and using
$\langle a_{k}(0)\rangle =\alpha _{k}$, we obtain the normal part
\begin{eqnarray}
E_{0}^{(+)}(x,t) &=&\frac{1}{2\pi }\int 2L_{m}\sqrt{\frac{ck}{2\epsilon V}}%
\frac{\Omega ^{2}}{\Omega ^{2}+G^{2}k}  \nonumber \\
&&\times e^{-f^{2}(k-k_{0})^{2}}e^{ik(x-ct)}dk,  \label{50}
\end{eqnarray}%
and the super-luminal\ and sub-luminal (or negative) parts
\begin{eqnarray}
E_{\pm }^{(+)}(x,t) &=&\frac{1}{2\pi }\int L_{me}\sqrt{\frac{ck}{2\epsilon V}%
}\frac{G^{2}k}{\Omega ^{2}+G^{2}k}  \nonumber \\
&&\times e^{-f^{2}(k-k_{0})^{2}}e^{ikx-it(kc\pm \Theta _{k})}dk,
\end{eqnarray}%
where $L_{me}$ is the scale of integration along the propagation direction ($%
x$-axis) of the probe light. We then expand the terms $\sqrt{k},$ $\frac{%
\Omega ^{2}}{\Omega ^{2}+G^{2}k},$ $\frac{G^{2}k}{\Omega ^{2}+G^{2}k}$ and $%
\Theta _{k}$ around the central value $k_{0}$ of the wave vector of the
input probe light. We can neglect the higher-order terms of $\Delta $ $%
=k-k_{0}$ since $k$ is very close to $k_{0}$. Then, with the defined
parameters $A=G^{2}/\Omega ^{2}$, $\Omega _{0}=\Omega \sqrt{1+Ak_{0}}$, the
approximate expressions of field components are obtained analytically as
\begin{eqnarray}
E_{\pm }^{(+)}(x,t) &\approx &\frac{Ak_{0}\sqrt{\pi k_{0}}}{2f^{3}(1+Ak_{0})}%
\{2f^{2}+iD_{0}[x-(c\pm E_{0})t]\}  \nonumber \\
&&\times \exp [-\frac{1}{4f^{2}}[x-(c\pm E_{0})t]^{2}]e^{ik_{0}(x-c_{\pm }t)}
\label{e1}
\end{eqnarray}%
and
\begin{eqnarray}
E_{0}^{(+)}(x,t) &\approx &\frac{\sqrt{\pi k_{0}}}{2f^{3}(1+Ak_{0})}%
[2f^{2}+iF_{0}(x-ct)]  \nonumber \\
&&\times \exp [-\frac{1}{4f^{2}}(x-ct)^{2}]e^{ik_{0}(x-ct)},  \label{e0}
\end{eqnarray}%
where $c_{\pm }=c\pm \frac{\Omega _{0}}{k_{0}}$ and%
\begin{eqnarray*}
D_{0} &=&\frac{3+Ak_{0}}{2k_{0}(1+Ak_{0})}, \\
E_{0} &=&\frac{\Omega _{0}A}{2(1+Ak_{0})}, \\
F_{0} &=&\frac{1-Ak_{0}}{2k_{0}(1+Ak_{0})}.
\end{eqnarray*}%
The above analytic result proves again the coexistence phenomena of both the
super-luminal\ and sub-luminal (or negative) group velocities. From the Eqs.
(\ref{e1},\ref{e0}), it's obvious that the group velocity of the part $%
E_{0}^{(+)}(x,t)$ is $V_{g_{0}}=c,$ and the group velocity of wave packet $%
E_{\pm }^{+}(x,t)$ is
\begin{equation}
V_{g_{\pm }}=c\pm E_{0}=c(1\pm \frac{\Omega }{2k_{0}c}\frac{n}{\sqrt{1+n}})
\end{equation}%
which is the same as the Eqs. (\ref{48}) and (\ref{49}).

Through the above theoretical analysis based on dynamic algebraic method,
several interesting properties of the light pulse propagation in an EIT
medium are observed. The central result is the possibility of coexistence of
"sub-luminal" and "super-luminal"\ group velocity phenomena, occurring as a
triple split of wave packet when spreading in such an EIT medium. We also
observe that if we inject coherent light into the EIT medium for a certain
time, there may appear three light pulses. One of these three light pulses
still propagates at the velocity $c$, but the other two have\
"super-luminal"\ and "sub-luminal" group velocities respectively.\ The
"sub-luminal" group velocity might even have negative value in some case.
This result is different from that of stable process in which only one
single group velocity appears. This is because we here only consider the
transient phenomenon or property. It is also remarked that both
"super-luminal"\ and \textquotedblleft sub-luminal" (even or negative) group
velocities are natural properties of light \cite{super} and could be
realized experimentally.

\section{Numerical Simulation of Wave Packet Split}

In this section we numerically simulate the dynamic process of the light
pulse propagating in the above described EIT medium. According to the Eqs. (%
\ref{e1},\ref{e0}), we calculate the shape evolutions of the wave packets of
the three parts of the light field, respectively. Let $\lambda _{0}=\frac{%
2\pi }{k_{0}}=\frac{2\pi c}{\omega _{0}}$ be the central wavelength of the
initial wave packet of the light pulse. To simulate the light pulse
propagating in the EIT medium numerically we distinguish the following two
situations: (I) the spatial width$\ d$ of light pulse is less than $\lambda
_{0}$; (II) $d$ contains a few $\lambda _{0}$'s.\ In the second situation\
several oscillations will be clearly observed in a light pulse with a few
frequency components. Let $E_{i}(x,t)=E_{i}^{+}(x,t)+$ $E_{i}^{-}(x,t)$, for
$i=0,+,-$. For situation (I) that the spatial width$\ d$ of light pulse is
less than its central wavelength ($d\approx 0.3\lambda _{0})$, three
3-dimension curves of $E_{0}(x,t)$, $E_{+}(x,t)$ and $E_{-}(x,t)$ are
plotted as in Fig. 2. It is observed from Fig. 2 the three parts $E_{0}(x,t)$%
, $E_{+}(x,t)$ and $E_{-}(x,t)$ of light field indeed have different group
velocities. It's noted that in order to satisfy the near-resonance
condition, the experimental probe light pluse contains large numbers of
wavelengths much more than that given above where it's convenient to see the
pulse splitting.

\begin{figure}[h]
\begin{center}
\includegraphics[width=6cm,height=11cm]{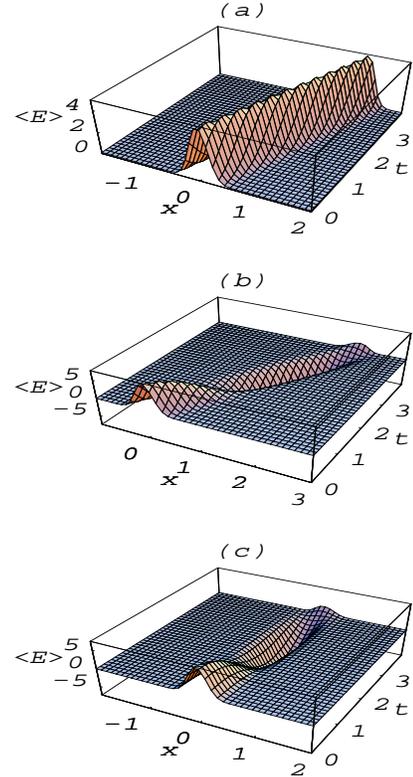}
\end{center}
\caption{ Wave packet split of light pulse in the case that the spatial
width of light pulse is less than its central wavelength $d\approx 0.3%
\protect\lambda _{0})$: (a): the normal part with group velocity $c$. (b):
the super-luminal part with group velocity being about $1.75c$. (c) the
sub-luminal part with group velocity being about$0.25c$.}
\end{figure}

Fig. 3 is drawn \ for the situation (II) that the spatial width$\ d$ of
light pulse is large than its central wavelength ( $d\approx 4\lambda _{0})$%
. In both of the situations it is observed that the group velocity ($v\simeq
1.75c)$ of $E_{+}(x,t)$ is larger than that of $E_{0}(x,t),$ but the group
velocity ($v\simeq 0.25c)$ of $E_{-}(x,t)$ is less than that of $E_{0}(x,t).$%
The only difference between these two situations is the details of the
oscillations of the caves.
\begin{figure}[h]
\begin{center}
\includegraphics[width=6cm,height=11cm]{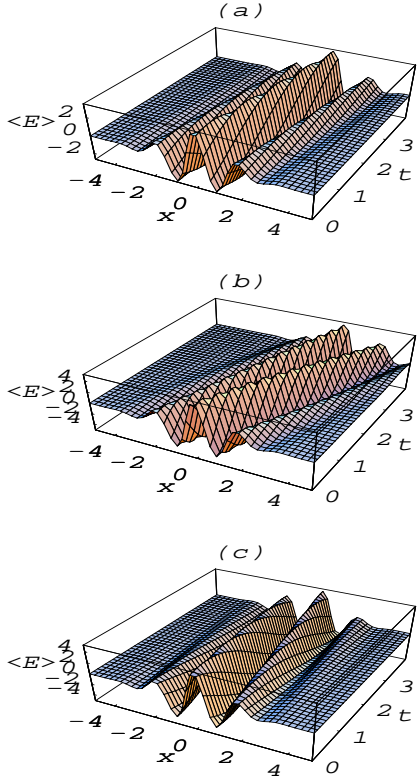}
\end{center}
\caption{ Wave packet split of light pulse in the case that the spatial
width of light pulse is large than its central wavelength $d\approx4\protect%
\lambda _{0})$: (a): the normal part with group velocity $c$. (b):the
super-luminal part with group velocity $v=0.25c$. (c): a sub-luminal part
with group velocity $v=1.75c$. The other parameters is as the same as that
in Fig. 2.}
\end{figure}

When the system parameters are changed so that
\begin{equation}
\frac{\Omega }{2k_{0}c}\frac{n}{\sqrt{1+n}}>1.
\end{equation}%
a seem-to-be exotic phenomenon of light pulse propagation is illustrated in
Fig. 4. Here, while two parts of the splitting wave packet (Fig.4a-4b)
possess normal group velocity $c$ and "super-luminal"\ group velocity $%
(v\simeq 2.36c)$ respectively, the third part propagates at a negative group
velocity $(v\simeq -0.36c).$


\begin{figure}[h]
\begin{center}
\includegraphics[width=6cm,height=11cm]{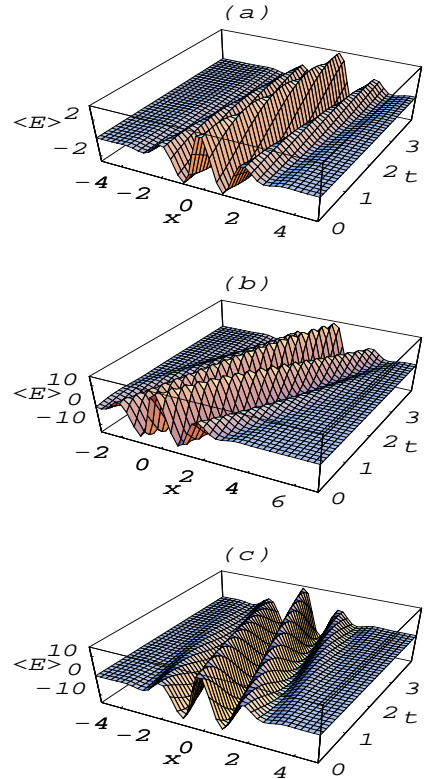}
\end{center}
\caption{ Propagation of light pulse with a part possessing a negative group
velocity $(\simeq -0.36c)$. Here, the spatial width of light pulse is large
than its central wavelength $d\approx 4\protect\lambda _0)$: (a): the normal
part with group velocity $c$. (b) the super-luminal part with group velocity
$v\simeq-0.36c$. (c) the part propagating with a negative group velocity $%
(v\simeq 2.36c).$ }
\end{figure}
\bigskip

In fact the amplitude of these field components depends on the system
parameters.\ It is easily seen from Eq. (\ref{45}) that the intensity of $%
E_{+}(x,t)$ is equal to that of $E_{-}(x,t)$, but is not equal to that of $%
E_{0}(x,t)$. The ratio of the amplitude of $E_{\pm }(x,t)$ to that of $%
E_{0}(x,t)$ is approximately
\begin{equation}
\frac{\sin ^{2}\theta _{k_{0}}}{2\cos ^{2}\theta _{k_{0}}}=\frac{%
g_{k_{0}}^{2}N}{2\Omega ^{2}}=\frac{n}{2}.
\end{equation}%
When the ratio $\frac{n}{2}$ is very small ($\ll 1$), most part of the light
pulse will propagate in the EIT medium at the group velocity $c$. If the
ratio $\frac{n}{2}$ is very large ($\gg 1$), then most part of the light
pulse propagates at "super-luminal"\ or sub-luminal (even negative) group
velocity. To see the dynamic details clearly we also plot the caves of the
spatial wave packets for different fixed instances in Fig. 5.

%
\begin{figure}[h]
\begin{center}
\includegraphics[width=9cm,height=9cm]{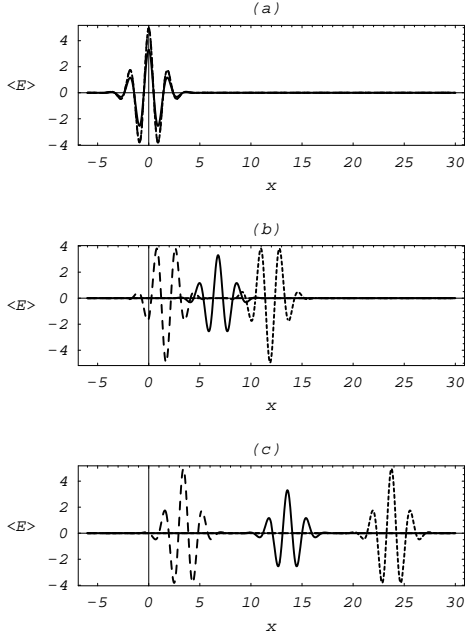}
\end{center}
\caption{Evolution of three splitting wave packets of light pulse. Here, $d=4%
\protect\lambda _{0})$ (a) The initial un-split wave packet with the spatial
width of light pulse $d\approx4\protect\lambda _{0})$;(b) [(c)] The shapes
of the three splitting wave packets at times $t_1\approx 2.3*10^{-13} s$ [ $%
t_2\approx 4.6*10^{-13}s$]. The cases with normal, "super-luminal" and
sub-luminal are depicted respectively by solid line,dot line and dash line.
The parameters are the same as that in Fig. 3}
\end{figure}

Fig. 5 (a-c) shows the time evolution of these three wave-packets at $%
t=t_{0}(=0),t_{1},t_{2}$ ($0<t_{1}<t_{2}$) according to the Fig. 3. The
field components $E_{+}(x,t)$ and $E_{-}(x,t)$) with "super-luminal"\ and
"sub-luminal" group velocities change their shapes of wave-packet during the
evolution, but the part $E_{0}(x,t)$ always propagates at $c$ \ with an
unchanged shape of wave-packet. Due to the intrinsic quantum coherence\ the
EIT medium does not change the height of each wave-packet. Strictly
speaking, $c=c_{0}/n_{0}$ is not the vacuum velocity of light, while $c_{0}$
is,where $n_{0}$ is the index of refraction \cite{sargent} (here for
simplicity, we set $n_{0}=1$ in our numerical simulation in this work). To
see the exotic phenomenon with negative group velocity, we take the
2-dimension curves at certain time in Fig. 6 according to the Fig. 4. Here,
a inverse-direction "light propagation"\ can be seen clearly.

%
\begin{figure}[h]
\begin{center}
\includegraphics[width=9cm,height=9cm]{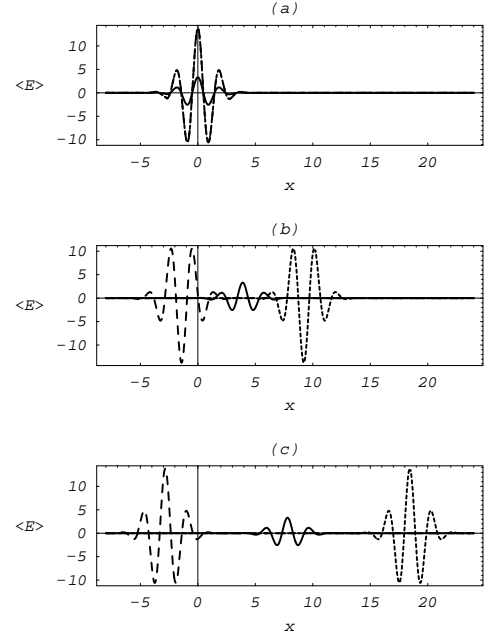}
\end{center}
\caption{ Phenomenon with a negative group velocity $(v\simeq -0.36c)$.
Here, $d=4\protect\lambda _{0})$: (a) The initial un-split wave packet with
the spatial width of light pulse $d\approx 4\protect\lambda _{0})$;(b) [(c)]
The shapes of the three splitting wave packets at times $t_{1}\approx
1.3\ast 10^{-13}s$ [ $t_{2}\approx 2.6\ast 10^{-13}s$]. The cases with
normal and super-luminal are depicted respectively by solid line and dot
line. The dash line describe the a part of the split wave packet propagating
with a a negative group velocity $(v\simeq -0.36c).$ The parameters are the
same as that in Fig. 4.}
\end{figure}

We can also calculate the optical intensity in the case, where the initial
quantized light field is multi-mode coherent state. We have
\begin{eqnarray}
I(x,t) &=&\left\langle E^{-}(x,t)E^{+}(x,t)\right\rangle   \nonumber \\
&=&E_{0}^{+}E_{0}^{-}+E_{+}^{+}E_{+}^{-}+E_{-}^{+}E_{-}^{-}  \nonumber \\
&&+2%
\mathop{\rm Re}%
(E_{0}^{+}E_{+}^{-}+E_{+}^{+}E_{-}^{-}+E_{-}^{+}E_{0}^{-}).
\end{eqnarray}%
From the Eqs. (\ref{e1}) and (\ref{e0}), it's obvious that $%
E_{0}^{+}E_{0}^{-}$, $E_{+}^{+}E_{+}^{-}$ and $E_{-}^{+}E_{-}^{-}$ are three
approximate Gaussian wave packets respectively and $E_{0}^{+}E_{+}^{-}$
(also $E_{+}^{+}E_{-}^{-}$ and $E_{-}^{+}E_{0}^{-}$) is the interference
term. At the initial time $t=0$, $I(x,t)$\ is a Gaussian wave packet
approximately since the center of each term contributing to $I(x,t)$ is at $%
x=0$. But after a certain time, the terms $E_{0}^{+}E_{0}^{-}$, $%
E_{+}^{+}E_{+}^{-}$ and $E_{-}^{+}E_{-}^{-}$ are nearly separated in spatial
coordinate and the contribution of interference terms will be close to zero.
So in this case the intensity will only contain 3 separated Gaussian wave
packets with the "super-luminal", "sub- (even negative) luminal" and normal
group velocities respectively.

Finally we consider the quantum fluctuation of the probe quantum field. In
the above discussions we assume the quantized light field of each mode\ to
be initially prepared in its coherent state. This is too conceptual a setup
for experiment. If the initial state is a number state the expectation of
the quantized light field operator vanishes for the representation of photon
number conservation. For this reason we need to consider the quantum
fluctuation described by the correlation function

\begin{eqnarray}
&&\left\langle E^{-}(x,t)E^{+}(x,t+\tau )\right\rangle   \nonumber \\
&=&\sum_{k}\frac{\hbar c\left\vert k\right\vert \left\langle n\right\rangle
_{k}}{2\epsilon V}e^{ikc\tau }(\cos ^{2}\theta _{k}+\sin ^{2}\theta _{k}\cos
\Theta _{k}t)^{2}.
\end{eqnarray}%
It will determines the intensity spectra
\begin{equation}
S(x,\omega )\equiv \int_{-\infty }^{\infty }\!\!{\rm d}\tau \,{\rm e}%
^{-i\omega \tau }\,\left\langle E^{-}(x,t)E^{+}(x,t+\tau )\right\rangle
=S(0,\omega ).
\end{equation}%
The one-time correlation is substantially the intensity of the probe light
field
\begin{eqnarray}
\left\langle I(x,t)\right\rangle  &=&\left\langle
E^{-}(x,t)E^{+}(x,t)\right\rangle   \nonumber \\
&=&\sum_{k}\frac{\hbar c\left\vert k\right\vert \left\langle n\right\rangle
_{k}}{2\epsilon V}(\cos ^{2}\theta _{k}+\sin ^{2}\theta _{k}\cos \Theta
_{k}t)^{2}.
\end{eqnarray}%
As a modulated asymptotic function, this result is illustrated in Fig. 7 \
for an initial number state with $\left\langle n\right\rangle _{k}$
satisfying a Gaussian wave distribution in frequency domain. It shows the
instantaneous process of light propagation to approach a stable state. For
very large Rabi coupling constant $\Omega ,$ $\theta _{k}\simeq 0,$ $\Theta
_{k}=\sqrt{g_{k}^{2}N+\Omega ^{2}}\simeq \Omega $ and the \ intensity
approaches a constant $\left\langle I(x,t)\right\rangle =\sum_{k}\frac{\hbar
c\left\vert k\right\vert \left\langle n\right\rangle _{k}}{2\epsilon V}.$ On
the other hand for a very large medium enhanced coupling of probe light $%
g_{k}^{2}N$ or $\Omega =0,\theta _{k}\simeq \frac{\pi }{2},$we can
analytically show
\begin{eqnarray}
\left\langle I(x,t)\right\rangle  &=&\sum_{k}\frac{\hbar c\left\vert
k\right\vert \left\langle n\right\rangle _{k}}{2\epsilon V}(\cos
g_{k}^{2}Nt)^{2}  \nonumber \\
&=&\sum_{k}\frac{\hbar c\left\vert k\right\vert \left\langle n\right\rangle
_{k}}{2\epsilon V}\cos ^{2}(\frac{c\wp ^{2}N}{2\hbar \epsilon V}kt),
\end{eqnarray}%
which is asymptotic to a constant with the modulation frequency $\omega _{M}=%
\frac{c\wp ^{2}N}{2\hbar \epsilon V}k$.

%
\begin{figure}[h]
\begin{center}
\includegraphics[width=6cm,height=5cm]{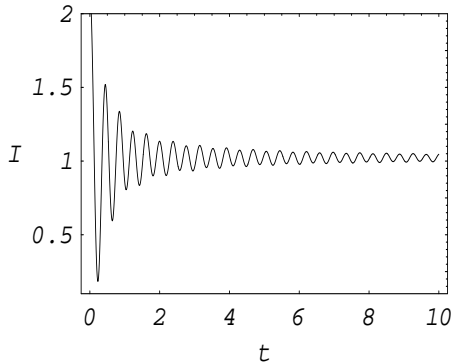}
\end{center}
\caption{The intensity evolution of probe light with a modulated asymptotic
behavior.}
\end{figure}

\bigskip

\section{Conclusion with Remarks}

In conclusion, we have investigated how the quantized light propagates in a
three-level $\Lambda -$type EIT medium by directly solving the Heisenberg
evolution equation of this light field based on the dynamic group method.
For simplicity, the quantized light field is considered as resonantly
coupling to the $\Lambda -$type subsystem even though this light field
contains a series of light modes and can not be resonant simultaneously. The
\ physical reason for this consideration is that the light wave packet has a
small width in frequency domain. In fact, the quantized light is considered
as a quasi-plane wave and propagating in the medium without the boundary
effect. It should be noticed that the present treatment is valid only when
the EIT medium is prepared in the low density excitation situation.

We wish to emphasize again that it is also owing to the wave nature of light
that the coexistence phenomena of both the "super-luminal"\ and
"sub-luminal" (or negative) group velocities appears as predicted in this
paper. It is very interesting to observe this coexistence phenomena
experimentally and explore its potential application in quantum memory and
quantum information process. To this end we need more details of physical
considerations on the experimental techniques. For instance, we need to
compare the size of the sample of the EIT medium and the split distances of
three wave packets resulting from the light pulse. In principle due to the
different group velocities the evolution of sufficiently-long time will
distinguish the wave packets, which might not preserve their shape because
of wave packet spreading or the dissipation and decoherence due to the
coupling to environment. On the other hand, what we predict are only the
transient phenomena. \ So it is somehow difficult to observe this
coexistence phenomena of both the "super-luminal"\ and "sub-luminal" (or
negative) group velocities in a practical experiment.

{\it This work is supported by the NSF of China (CNSF grant No.90203018) and
the knowledged Innovation Program (KIP) of the Chinese Academy of Science.
It is also founded by the National Fundamental Research Program of China
with No 001GB309310. We also sincerely thank Y. Wu, P. Zhang and L. You for
the useful discussions with them.}

\thispagestyle{myheadings} \markright{Sub-dynamics } \appendix

\section*{{\bf {Light Propagation in Exciton System: Analytic Result from
Sub-dynamics Mehtod }}}

\renewcommand{\theequation}{A.\arabic{equation}} \setcounter{equation}{0} %
\vskip 0.511cm \renewcommand{\theequation}{A.\arabic{equation}} %
\setcounter{equation}{0} \vskip0.511cm

In general we consider a physical system with a dynamic symmetry
characterized by a Lie group $G$. This means that the Hamiltonian of the
considered system
\begin{equation}
H=H[G]\equiv H(g_{1},g_{2},...)
\end{equation}%
is a functional of the generators $g_{1},$ $g_{2},...$ of $G$. These
generators \ can be understood as the basic dynamic variables of the system.
Suppose there exists a subgroup $S\subset G$ such that
\begin{equation}
\lbrack H,G]\subset S.
\end{equation}%
Let $s_{1},$ $s_{2},...$ be the generators of $S$. Then the system of the
Heisenberg equations
\begin{equation}
i\frac{d}{dt}s_{k}=[s_{k},H]\in S
\end{equation}%
about $s_{1},$ $s_{2},...$ are closed to form the so-called sub-dynamics.
Through the subset $\{s_{1},s_{2},...\}$ of the complete set of dynamic
variables $g_{1},$ $g_{2},...$, the sub-dynamics depict the main features of
the considered system.

The excitonic system in this paper serves as a practical example for
\begin{equation}
G=SU(2)\stackrel{-}{\otimes }\Xi \otimes \Gamma
\end{equation}%
and $S=$ $\Xi \otimes \Gamma $ where $\Gamma $ is the Heisenber-Weyl group
generated by the creation and annihilation operators $a_{k}^{\dagger }$ and $%
a_{k}$ of light field.


\end{document}